# Chromium-Doped Bismuth Antimony Telluride for Future Quantum Hall Resistance Standards

**Albert F. Rigosi**

Linsey K. Rodenbach, Alireza R. Panna, Shamith U. Payagala, Ilan T. Rosen, Joseph A. Hagmann, Peng Zhang, Lixuan Tai, Kang L. Wang, Dean G. Jarrett, Randolph E. Elmquist, Jason M. Underwood, David B. Newell, and David Goldhaber-Gordon

*National Institute of Standards and Technology (NIST), Gaithersburg, MD 20899, United States*

afr@nist.gov

Abstract: Since 2017, epitaxial graphene has been the base material for the US national standard for resistance. A future avenue of research within electrical metrology is to remove the need for strong magnetic fields, as is currently the case for devices exhibiting the quantum Hall effect. The quantum Hall effect is just one of many research endeavours that revolve around recent quantum physical phenomena like composite fermions, charge density waves, and topological properties [1-2]. New materials, like magnetically doped topological insulators (MTIs), offer access to the quantum anomalous Hall effect, which in its ideal form, could become a future resistance standard needing only a small permanent magnet to activate a quantized resistance value [3-5]. Furthermore, these devices could operate at zero-field for measurements, making the dissemination of the ohm more economical and portable. Here we present results on precision measurements of the $h/e^2$ quantized plateau of Cr-Doped $(Bi_xSb_{1-x})_2Te_3$ and give them context by comparing them to modern graphene-based resistance standards. Ultimately, MTI-based devices could be combined in a single system with magnetic-field-averse Josephson voltage standards to obtain an alternative quantum current standard.

Text: For resistance metrology, this handoff manifested itself as the development of graphene-based technology, with devices capable of accessing a quantized Hall resistance (QHR). Realizing the unit of the ohm with graphene-based devices, once an exclusive accomplishment and capability for National Metrology Institutes (NMIs) and primary standards laboratories [3] – [7], has become more user-friendly due to simplification of equipment infrastructure and the longer shelf-life for improved devices with large-scale crystallinity [8] – [11].

Despite its popularity, graphene-based devices have been found to have some limitations regarding their operability and accessible parameter space. The most evident requirement that adds additional equipment and complexity is that of a strong magnetic field needed for the quantum Hall effect to be exhibited in a metrologically robust way. Newer materials, like magnetically doped topological insulators (MTIs), offer access to the quantum anomalous Hall effect (QAHE), which is a phenomenon linked to the breaking of time-reversal symmetry and the opening of an energy gap that can accommodate the existence of topological surface states [12] – [14].

Under typical circumstances, this formed energy gap closes where the component of the magnetization normal to the surface changes direction. For samples that approach the two-dimensional limit and have uniform out-of-plane magnetization, such a transition would occur at the edge of the sample since this characteristic vector switches direction as it goes from the top to the bottom surface. This switching leaves one-dimensional channels that are chiral and dissipationless due to the absence of available states for backscattering. In its ideal form, the QAHE could become the basis for a future resistance standard because it would need only a small permanent magnet to activate a quantized resistance value. Furthermore, these devices could operate at zero-field for measurements, making the dissemination of the ohm more economical and portable.

This work presents precision measurement results of the $h/e^2$ quantized plateau of Cr-Doped $(Bi_xSb_{1-x})_2Te_3$ and gives them context by comparing them to other recent results. Measurement techniques and examinations of the QAHE are also presented to prime the community for this potential paradigm shift, as efforts revolving around using MTIs for metrology continue to grow. Additionally, this forward-looking study will examine typical critical currents, longitudinal and Hall resistivity behaviors, and measurement noise data to better understand how one should expect MTI-based devices to be used as possible successors to graphene-based technology (both single-element and p-n junction-based devices [15]). Accompanying uncertainty budgets are described and show that MTIs are within the realm of being applied as standards in the near future, despite stringent temperature requirements.

The growth of the 6 quintuple layers of high-quality single-crystalline Cr-Doped $(Bi_xSb_{1-x})_2Te_3$ film took place on a semi-insulating GaAs (111)B substrate, with details provided in other work [12]. MTI-based devices were fabricated using direct-write optical photolithography. The electrical contact pads consisted of a 5 nm Ti adhesion layer and 100 nm Au. To implement a top gate, a dielectric layer of 1 nm of Al had to first be uniformly deposited to act as a seed layer. This thin layer was oxidized before depositing about 40 nm of Al. The top gate was deposited by evaporating 5 nm Ti and 85 nm Au.

Devices were mounted onto a chip carrier (JCC2824001 package, or JCC-28), which was tested to ensure that it was non-magnetic. An example diagram of the sample design and optical image are shown in Fig. 1 (a) and (b). It should be noted that there are two pairs of orthogonal electrical contacts for measuring the QAHE (labeled $\rho_{xy}$ and $\rho'_{xy}$) and two pairs of contacts for measuring the longitudinal resistivity.

For precision measurements of the device, a 12-bit binary cryogenic current comparator (CCC) was used. A CCC realizes an unknown resistance ratio $R_1/R_2$, from the inverse ratio of the dc bias currents $I_1$ and $I_2$. It consists of a superconducting toroidal screen which houses the superconducting windings $N_1$ and $N_2$. When a dc current is passed through these windings, due to the Meissner effect, a net current exists on the surface of the superconducting screen. The magnetic flux due to this net current imbalance is detected using a superconducting quantum interference device.

The assessment of the MTI-based devices is shown in Fig. 2. A summary of the precision measurements of $\rho_{xx}$ and the QAHE are shown in (a) and (b), respectively, with data also including two summarizing points from previous studies. Applied currents vary and the following MTI measurements are represented: 2018 NIST/Stanford University (green), Physikalisch-Technische Bundesanstalt/University of Würzburg (red), this work's first device (black), and the second device (blue). The cyan region in (b) marks the boundary for data points going beneath one part in $10^7$. All error bars represent $k = 1$ type A and B combined uncertainties.

**Commercial equipment, instruments, and materials are identified in this paper in order to specify the experimental procedure adequately. Such identification is not intended to imply recommendation or endorsement by the National Institute of Standards and Technology or the United States government, nor is it intended to imply that the materials or equipment identified are necessarily the best available for the purpose.**

**Work presented herein was performed, for a subset of the authors, as part of their official duties for the United States Government. Funding is hence appropriated by the United States Congress directly. This material is based upon work partially supported by the Department of Energy, Office of Science, Basic Energy Sciences, Materials Sciences and Engineering Division, under Contract DE-AC02-76SF00515.**

Figures

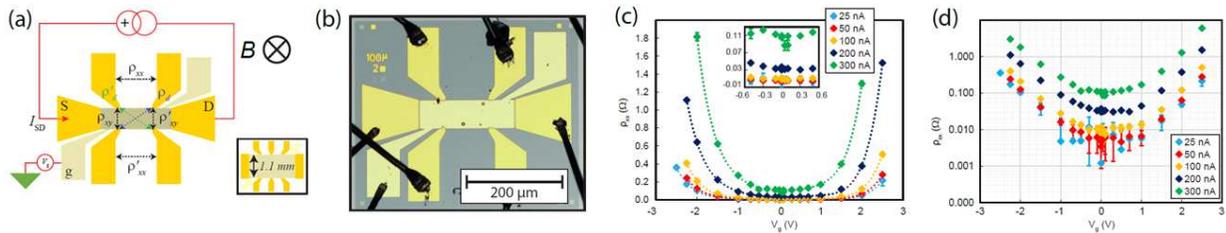

**Figure 1:** Fabrication of a MTI-based device and a basic top gate characterization of the longitudinal resistance.

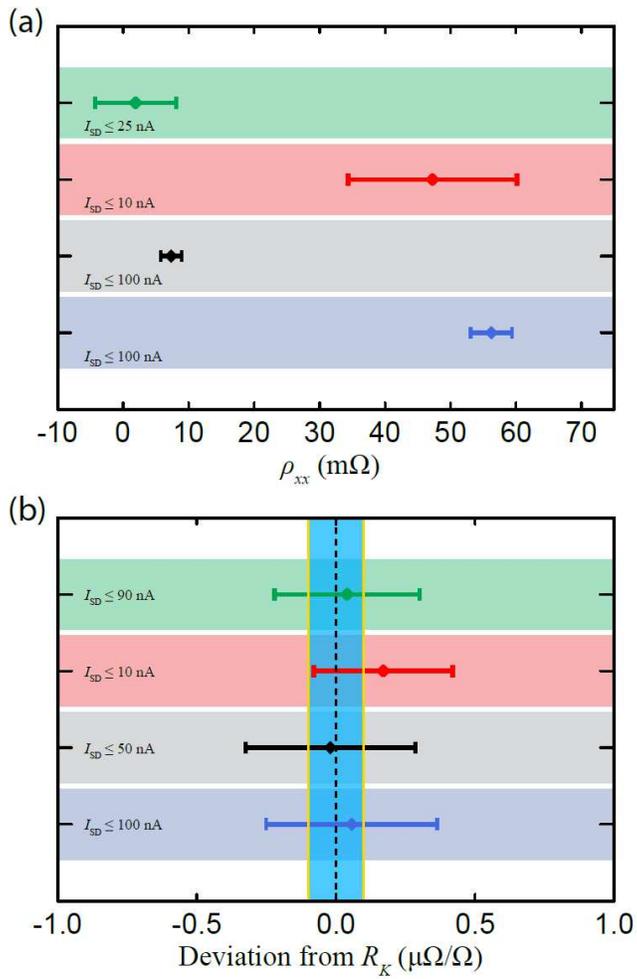

**Figure 2:** A summary of the precision measurements of (a) $\rho_{xx}$ and (b) the QAHE for the following MTI measurements are represented: 2018 NIST/Stanford (green), PTB/UW (red), this work's first device (black), and the second device (blue). The cyan region in (b) marks the boundary for data points going beneath one part in $10^7$. All error bars represent $k = 1$ type A and B combined uncertainties.